\newcommand{\CLASSINPUTbottomtextmargin}{2.54cm}
\newcommand{\CLASSINPUToutersidemargin}{2.6cm}
\newcommand{\CLASSINPUTinnersidemargin}{2.6cm}
\documentclass[conference]{IEEEtran}
\IEEEoverridecommandlockouts

\usepackage{xcolor}
\usepackage{subfigure}
\usepackage{color}
\usepackage{graphicx}
\usepackage{balance}
\usepackage{amsmath}
\usepackage{amsfonts}
\usepackage{amssymb}
\usepackage{pifont}
\usepackage{times}
\usepackage{paralist}
\usepackage{textcomp}
\usepackage[hyphens]{url}
\usepackage[font={footnotesize}]{caption}
\usepackage{units}
\usepackage{enumitem}
\usepackage{amsmath,colortbl}
\usepackage{cite,comment}
\usepackage[nolist]{acronym}
\usepackage{soul}
\usepackage{array}
\usepackage{tabularx}
\usepackage{multirow}
\usepackage{lipsum}
\begin{acronym}[ACRONYM]
  \acro{3GPP}{the third generation partnership project}
  \acro{4G}{fourth generation}
\acro{5G}{fifth generation}
\acro{6G}{sixth generation}
\acro{AI}{artificial intelligence}
\acro{AoA}{angle-of-arrival}
\acro{AoD}{angle-of-departure}
\acro{BS}{base station}
\acro{CDF}{cumulative density function}
\acro{MIMO}{multi-input multi-output}
\acro{IoT}{internet of things}
\acro{RSSI}{received signal strength indication}
\acro{CSI}{channel state information}
\acro{SV}{Saleh-Valenzuela}
\acro{ULA}{uniform linear array}
\acro{BS}{base stations}
\acro{CNN}{convolution neural network}
\acro{MFM}{Matlab for MIMO}
\acro{CDF}{cumulative distribution function}
\acro{RTI}{radio tomographic imaging}
\acro{JCAS}{joint communication and sensing}
\acro{ISAC}{integrated sensing and communication}
\acro{OFDM}{orthogonal frequency division multiplexing}
\acro{LOS}{line of sight}
\acro{NLOS}{non-line of sight}
\acro{DDP}{delay-Doppler profile}
\acro{PDP}{power delay profile}
\acro{RCS}{radar cross section}
\acro{SNR}{signal to noise ratio}
\acro{SNRs}{signal to noise ratios}
\acro{CPI}{coherent processing interval}
\acro{FD}{full duplex}
\acro{3GPP} {third generation partnership project}
\acro{CNN}{convolutional neural network}
\acro{CFAR}{constant false alarm rate}
\acro{PDF}{probability distribution function}
\acro{QPSK}{quadrature phase shift keying}
\end{acronym}

           \newcommand{\mbf}[1]{\mathbf{#1}}          
\newcommand{\mb}[1]{\mbox{#1}}

\begin{document}
\bstctlcite{IEEEexample:BSTcontrol}

\title{Indoor Sensing with Measurements}
\date{\today}
\author{Vijaya~Yajnanarayana, Philipp~Geuer, and Satyam~Dwivedi \\
Ericsson Research 
\thanks{This work was supported by the Federal Ministry of Education and Research Germany within the project “KOMSENS-6G”, grant no: 16KISK127.}
}
\maketitle

\begin{abstract}
The cellular wireless networks are evolving towards acquiring newer capabilities, such as sensing, which will support novel use cases and applications. Many of these require indoor sensing capabilities, which can be realized by exploiting the perturbation in the indoor channel.  In this work, we conduct an indoor channel measurement campaign to study these perturbations and develop AI-based algorithms for estimating sensing parameters. We develop several AI methods based on \acp{CNN} and tree-based ensemble architectures for sensing. We show that the presence of a passive target like a person can be detected from the channel perturbation of a single link with more than 90 \% accuracy with a simple \ac{CNN} based AI algorithm. However, sensing the position of a passive target is far more challenging requiring more complex AI algorithms and deployments. We show that the position of the human in the indoor room can be estimated within the average position error of 0.7\,m with a deployment having three links and employing complex AI architecture for position estimation. We also compare the results with the baseline algorithm to demonstrate the utility of the proposed method.
\end{abstract}


\section{Introduction}
\label{sec:intro}
The ability of a network to sense the target's presence and estimate its state parameters such as  position, velocity, trajectory, etc. provides an environmental awareness which is foreseen to be essential for many sensing use cases. Passive target sensing involves detecting unconnected targets, and estimating its state parameters.  In this paper, we focus only on the detection and position estimation of the passive target in a cluttered indoor environment.  Sensing extension to the ubiquitous communication infrastructure will not only resolve need for ambient lighting and cost-related concerns of vision based sensing systems, but will also extend the sensing coverage. 

\begin{figure*}[t]
  \centering
  \fbox{\includegraphics[width=1.6\columnwidth]{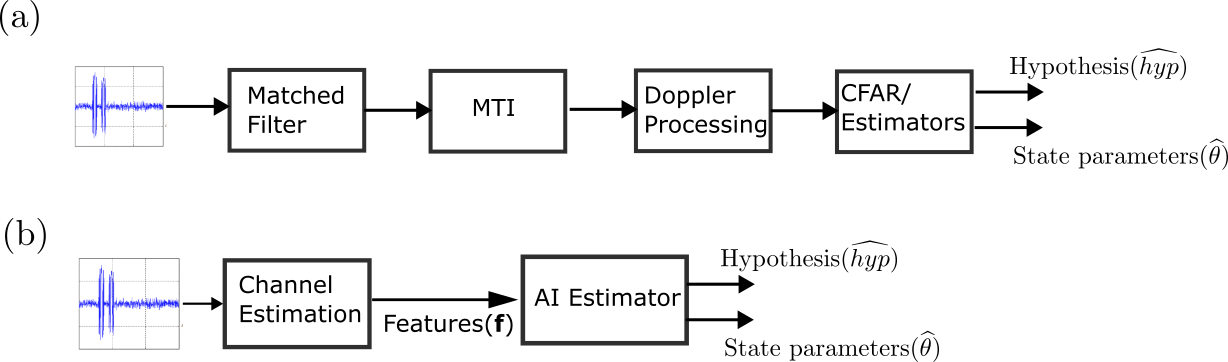}}
  \caption{Target sensing strategies: (a) Classical signal processing approach for target sensing. (b) A Hybrid method using channel estimation and AI for sensing. }
  \label{fig:intro} \vspace{-5mm}
\end{figure*}

Typical received sensing signals are impaired by clutter, noise, and interference. The high-level processing pipeline for sensing is shown in the Fig.~\ref{fig:intro}(a). Compared to signal processing approaches, AI methods can learn features automatically from data, this is favorable when encountered with complex environment with diverse target of interest. Though it is possible to use raw samples directly for sensing inference, we take a hybrid approach, in which we first estimate the channel using traditional methods and use the perturbations in them towards target sensing. This is further illustrated in Fig.~\ref{fig:intro}(b).

Numerous research studies have focused on indoor target detection and position estimation. In \cite{yajnanarayana2023eucnc}, authors model indoor channels using a stochastic geometric channel model called Saleh-Valenzuela (SV) model \cite{saleh-1987-statis-model}. They use a simulator discussed in \cite{mfm} to create a multi-static setup employing a sophisticated receiver with beamforming capabilities. This work exploits the fact that in high-frequency deployments, the presence of the target creates a shadow which blocks angle-of-arrivals (AoAs) within certain ranges to the receiver, thereby creating a perturbation in the channel state information (CSI).  Though AI methods proposed in the above work are promising, indoor channels are hard to model using stochastic models at high frequencies and hence deterministic channel models are favored. In \cite{yajnanarayana2024-bistatic}, authors use a deterministic channel model called scatter point channel model \cite{chris2023eucnc} with a bi-static deployment and use the delay-Doppler images for detecting targets. The current work is an extension of this work, except that we move away from a simplistic scatter point channel model to the real-world channel captured from the measurement campaign in a cluttered room to introduce practical clutter and realistic channel realizations.

\section{Measurement Campaign}
\label{sec:meascampaign}
\begin{figure}[t]
  \centering
  \includegraphics[width=0.8\columnwidth]{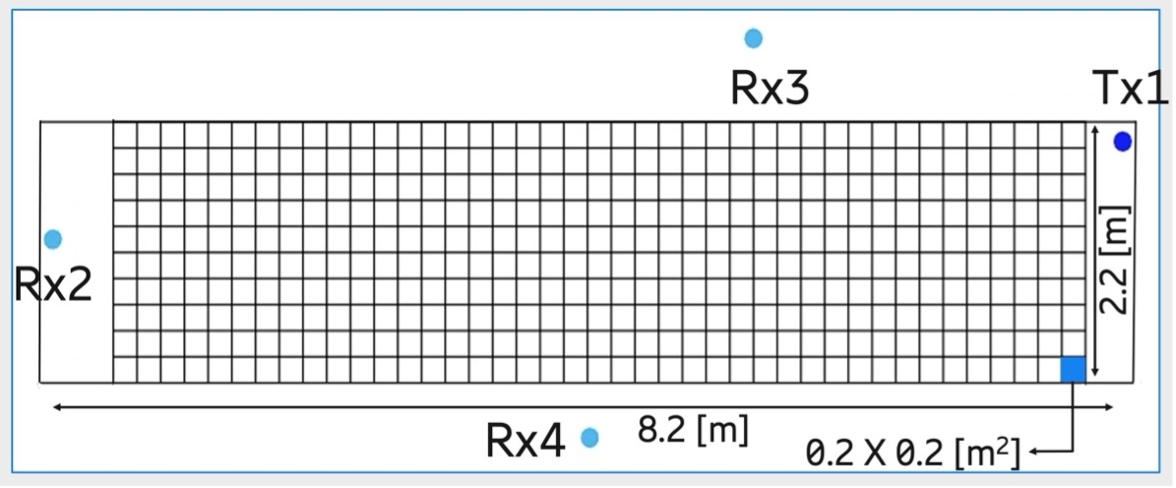}
  \caption{Measurement campaign included capturing the channel at 462 grid points between a single transmitter (Tx1) and three receivers (Rx2/3/4), with and without target.}
  \label{fig:grid}
\end{figure}

\subsection{Measurement Setup}
\label{ss:meassetup}
The measurement campaign was conducted in the propagation lab, Kista, Stockholm. The measurements were collected using Keysight PNA n5224a network analyzer. Fig.~\ref{fig:grid} shows the schematic of the grid laid on the floor of the lab. There were $462$ grid points with a cell size of $20\mb{\,cm} \mb{ x } 20 \mb{\,cm}$. As can be seen in the schematic, there is one transmitter and three receivers in the setup. The transmitter and the receivers are connected to the ports of the network analyzer.

\subsection{Measurement Analysis}
\label{ss:measanalysis}
The network analyzer sweeps $N$ frequency points, $f_i  | i\in\{1,\ldots,N\}$, over a bandwidth $B$,  around a chosen center frequency, $F_c$. The collection of observations at these $N$ frequency points (captured during a frequency sweep), over $B=1\mb{ GHz}$ and $F_c=28\mb{ GHz}$ is given by  $X=\{X_1,X_2,\ldots,X_N\}$. Each observed frequency sample is given by
\begin{equation}
  \label{eq:sample}
  X_i=\sum_{l=1}^L a_l \exp \left\{-j 2 \pi\left(f_i+d_l\right) \tau_l\right\}+w_i,
\end{equation}
where $f_i$ is a frequency in the sweeping set and $L$ indicates the number of paths. The parameters $d_l$, $\tau_l$ and $a_l$ denote Doppler frequency, delay, and gain of the $l$-th path respectively. The noise in the $i$-th sample is given by $w_i$.

The channel impulse response, which is fed to the learning algorithm is obtained by taking the inverse Fourier transform ($\mathcal{F}^{-1}$) of  $X$, and is given by

\begin{equation}
  \label{eq:ifft}
  \mbf{h}=\mathcal{F}^{-1}\{X\}.
\end{equation}

\begin{table}[t]
  \centering
  \begin{tabular}[c]{|l|l|}
  \hline   
    Center Frequency & $28\mb{\,GHz}$ \\
    Bandwidth & $1\mb{\,GHz}$ \\
    Number of Transmitter & 1 \\
    Number of Receivers & 3 \\
    Number of grid points & 462 \\
    Grid resolution & $0.2\mb{\,m}$ \\
   \hline                   
  \end{tabular}
  \caption{Measurement Configuration}
  \label{tab:measconfig}\vspace{-3mm}
\end{table}
The measurement configuration is listed in the Table~\ref{tab:measconfig}. Before collecting the measurements, the calibration of measurement equipment is done to remove any unknown excess delays in the receiver links.

The measurement procedure involves collecting channel impulse responses (CIRs) between the transmitter and three receivers by having a person standing at different grid points (alternative hypothesis) shown in  Fig.~\ref{fig:grid}. Everything else in the indoor environment is kept static for the duration of measurement collection. An equal number of null hypothesis measurements (person absent in the environment) are also collected. This measurement set comprising of CIRs from three links for null and alternate hypothesis is used to train and evaluate AI algorithms.

\section{AI Architectures and Algorithms}
\label{sec:arch}
Below we describe various AI architectures and algorithms, which uses the measurements from the measurement campaign described above.

\subsection{CNN Based Architectures}
\label{ss:archCNN}
\begin{figure}[t]
  \centering
  \fbox{\includegraphics[width=0.9\columnwidth]{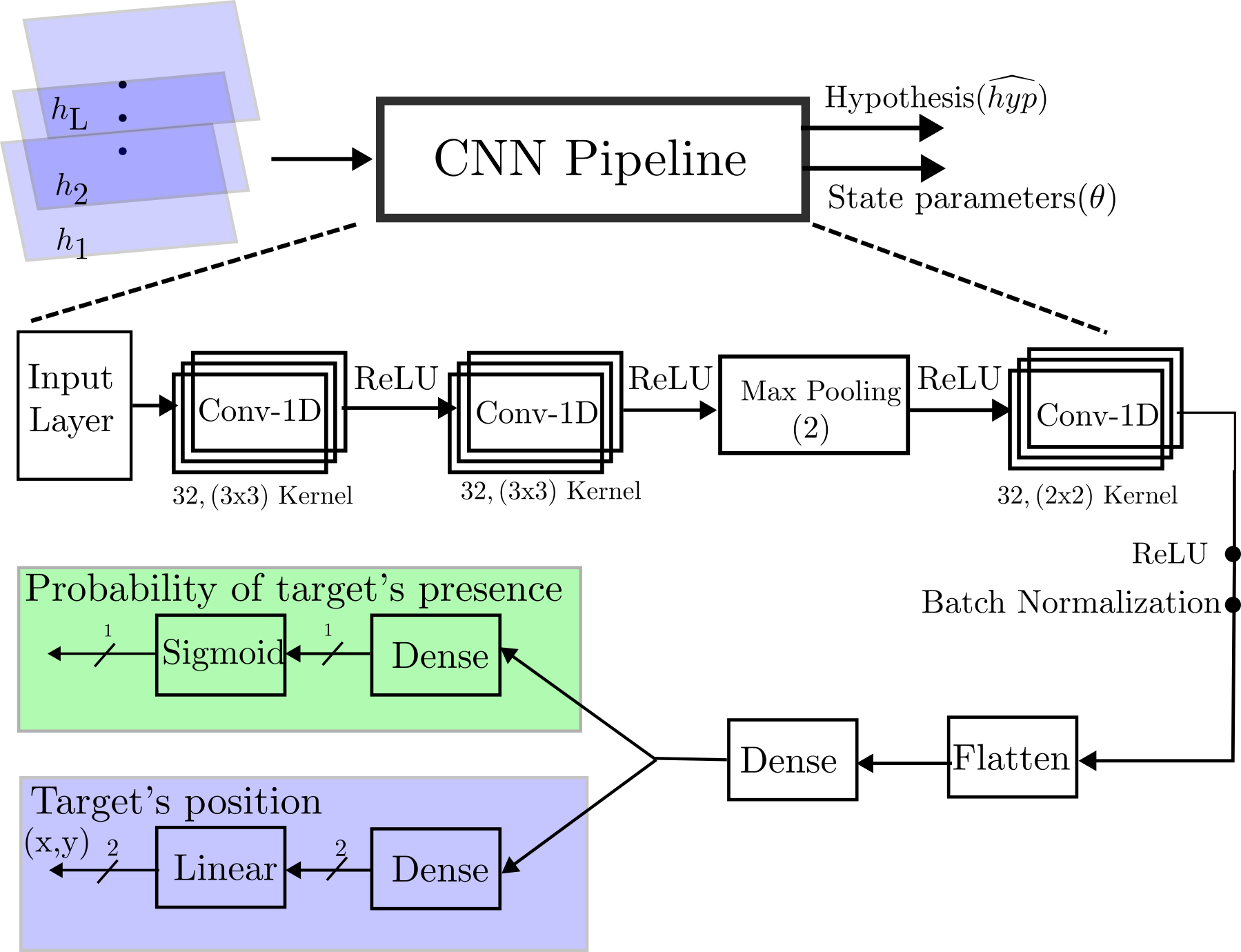}}
  \caption{Type-A AI architecture with CNN for target detection and position estimation.}
  \label{fig:typeA}
\end{figure}

In this study, two different CNN architectures are investigated. The first architecture, Type-A, is an extension of the shallow CNN structure discussed in \cite{yajnanarayana2024-bistatic}. This structure is illustrated in Fig~\ref{fig:typeA}. It comprises of a single CNN pipeline which can take multiple channels (defined based on the input signal dimension) into it. The pipeline consists of stacks of 1D convolution layers, non-linear activation function (ReLU) and pooling layers. 1D convolution layers with 32 kernels are used to extract spatial correlation in the input signal. The output of the convolution layer is passed through a non-linear activation (ReLU) for non-linear modeling and max pooling layer to enhance network's capability to generalize.

\begin{figure}[t]
  \centering
  \fbox{\includegraphics[width=0.9\columnwidth]{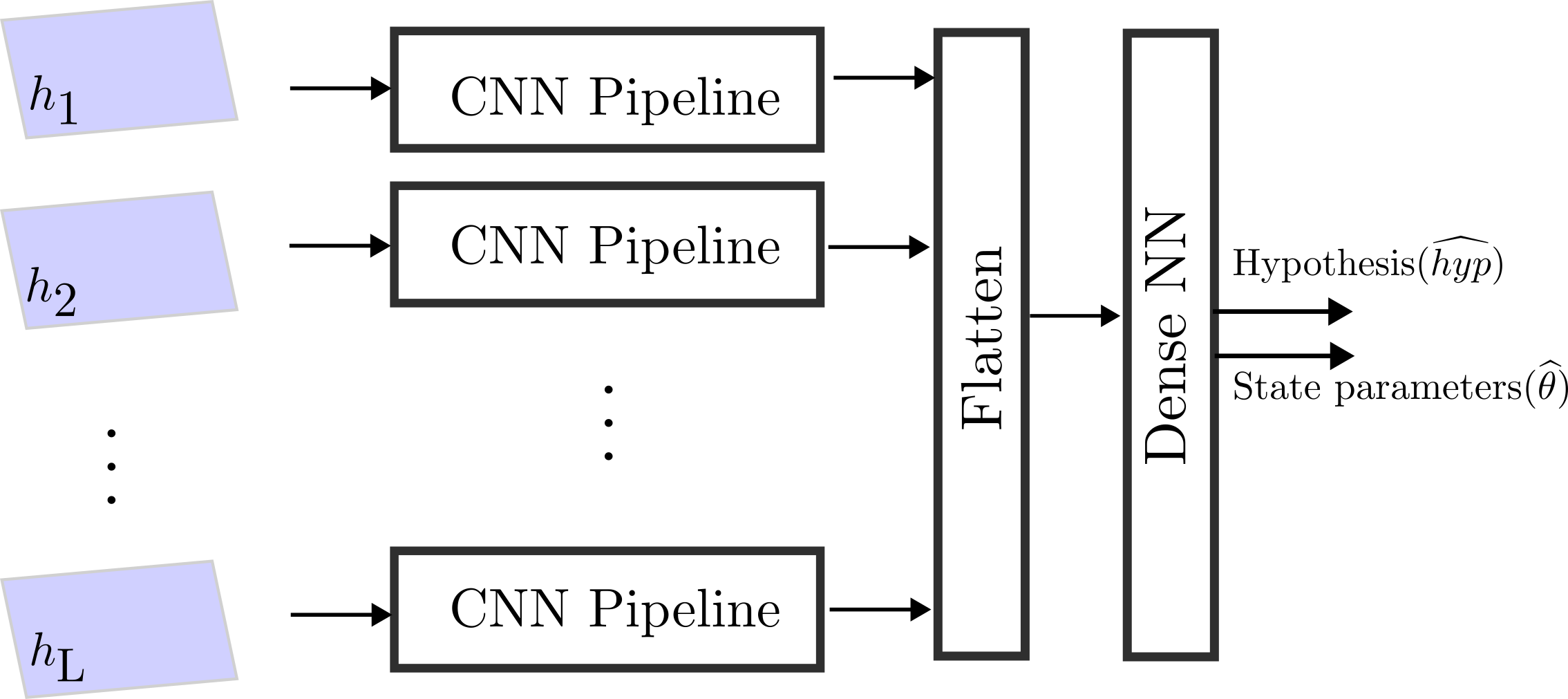}}
  \caption{Type-B AI architecture with CNN for position estimation.}
  \label{fig:typeB}
\end{figure}

The input layer comprises CIRs from $L$ links fed into different channels of the CNN pipeline described in Fig.~\ref{fig:typeA}. Since our deployment considered in the measurement campaign consisted of a single transmitter and 3 receivers, $L$ can take a maximum value of $3$.
We also studied a more complex CNN architecture (Type-B) with greater number of parameters and with improved ability to exploit patterns in the impulse response towards sensing. This is shown in Fig.~\ref{fig:typeB}. Here the impulse response from each link is fed into a different CNN pipeline and the features from them are fused using a dense network towards estimating sensing parameters.

\begin{figure}[t]
  \centering
  \fbox{\includegraphics[width=0.8\columnwidth]{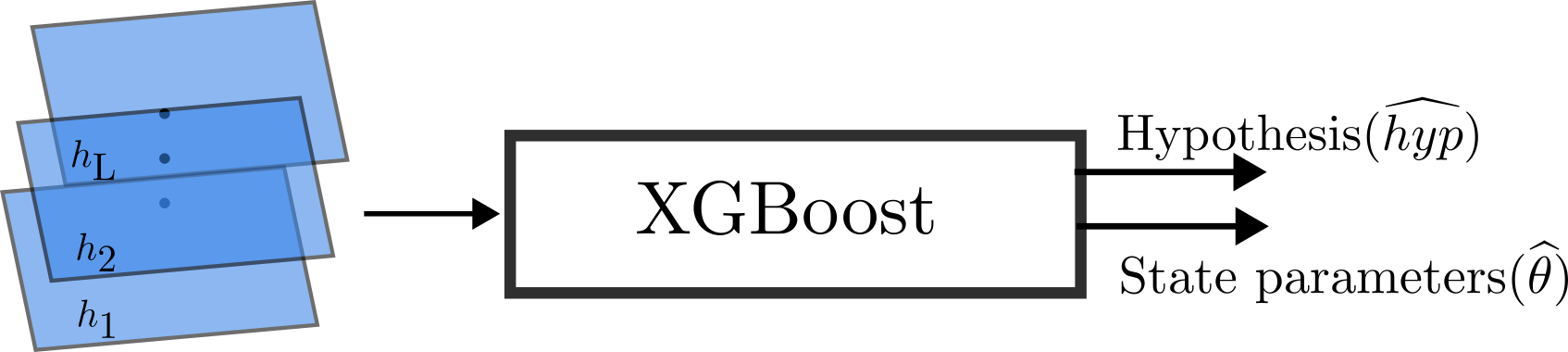}}
  \caption{Type-C AI architecture with XGBoost position estimation.}\vspace{-3mm}
  \label{fig:typeC}
\end{figure}

\subsection{Tree-Based Architectures}
\label{ss:archTREE}

Tree-based ensemble architectures are a class of machine learning algorithms that combine multiple decision trees to create a more robust and accurate predictive model. These algorithms generally operate by either aggregating the predictions of multiple trees in parallel (bagging) or sequentially refining the predictions of previous trees (boosting).

XGBoost (eXtreme Gradient Boosting) is a highly optimized and efficient implementation of the Gradient Boosting algorithm \cite{chen2016-xgboos}, which we consider for this work. XGB incorporates several advanced techniques and optimizations that enhance its performance and scalability. One of its key features is the use of a regularized objective function, which includes both a loss term, measuring the difference between predicted and actual values, and a regularization term, controlling the complexity of the model. The regularization term penalizes large leaf weights and deep trees, effectively preventing overfitting. XGB also supports parallel processing by multi-threading and distributed computing frameworks, enabling faster training times on large datasets. For that, tree splitting algorithm called “weighted quantile sketch,” is used which proposes candidate splitting points based on percentiles. Additionally, compared to Gradient Boosting XGB also uses tree pruning and can handle missing data.

The Type-C architecture using XGBoost is shown in Fig.~\ref{fig:typeC}. We employ hyperparameter tuning via grid-search to select parameters such as number of estimators, maximum depth and learning rate. Results show also that the Type-C architecture is less prone to overfitting than the CNN models and achieves a good generalization over a large range of lengths of channel impulse responses which will be discussed in next section.

\section{Results}
\label{sec:results}
\subsection{Target Detection Performance}
\label{ss:resultsTD}
Below we describe the performance of the proposed methods for hypothesis testing, that is, detecting the presence of the human in the scene. We consider the measurement set having channel impulse responses between transmitter Tx1 to receivers Rx2, Rx3 and Rx4 from 462 grid points (where a person is standing) for alternative hypotheses (i.e., target present) and an equal number of NULL hypotheses (i.e., target absent) as explained in Section~\ref{sec:meascampaign}. This measurement set is split exclusively into training and testing sets. Training set consists of impulse responses for $337$ random bins and testing set consists of impulse responses for $125$ random bins. Training set is further split  $70/30$ between training and validation, which is employed to train Type-A architecture shown in Fig.~\ref{fig:typeA}. The performance is assessed using the test set which is unseen by the trained agent.

\begin{figure}[t]
  \centering
  \fbox{\includegraphics[width=0.8\columnwidth]{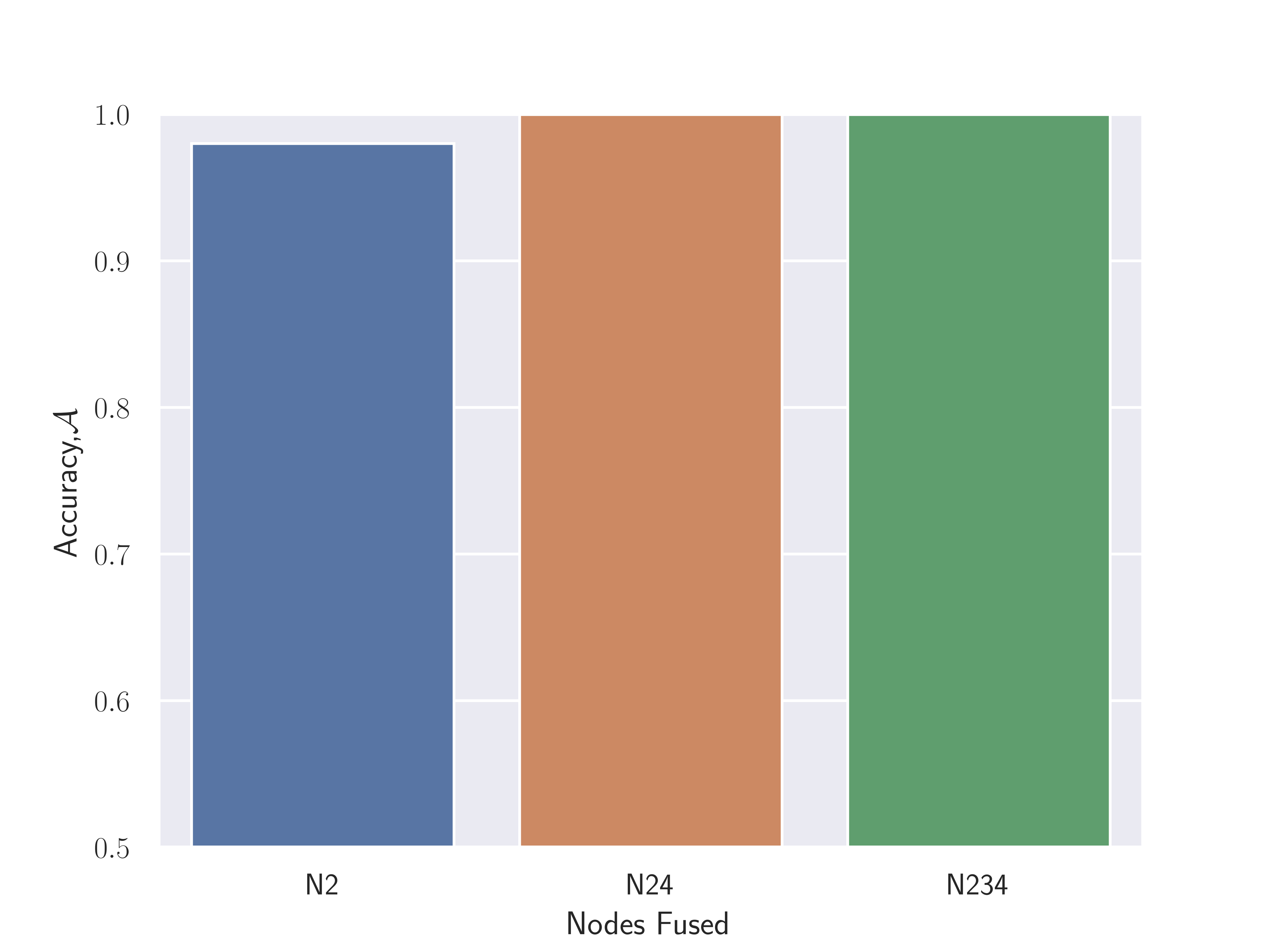}}
  \caption{Passive Target (human) detection performance using Type-A architecture.}
  \label{fig:detectionResult}
\end{figure}

The performance is reported in terms of accuracy and is shown in Fig~\ref{fig:detectionResult}. X-axis indicates the number of links involved, for example “N24”, corresponds to the performance of the algorithm using 2 links constituting of transmitter Tx1 to receiver Rx2, and transmitter Tx1 to receiver Rx4. As shown in Fig~\ref{fig:detectionResult}, even with the CIR from a single link and using the less complex Type-A algorithm, the passive target can be detected with an accuracy of more than 90\%. The performance can be further improved by exploiting the CIRs from multiple links. 

\subsection{Positioning Performance}
\label{ss:resultsPP}

\subsubsection{CNN Based Architectures (Type-A and Type-B)}
\label{sss:cnn}
We use the similar training and testing split explained in the previous section. However now the algorithm is trained to minimize the mean square error of the position. The performance of Type-A architecture is reported as CDF of the position error in Fig.~\ref{fig:PP_TypeABC}. The red line shows the performance when using CIR from a single link involving Transmitter Tx1 and Receiver Rx2, while the  green plot shows the performance when CIRs from all three links are used. When the CIRs from all three links are used, it will reduce the average position error from $\mu_{\varepsilon}^{2}=1.3\mb{\,m}$ to $\mu_{\varepsilon}^{234}=1.1\mb{\,m}$\footnote{The superscript in $\mu_{\varepsilon}$ indicates the receivers employed. For example, $\mu_{\varepsilon}^2$ indicates average position error considering only one link between Tx1 and Rx2, while $\mu_{\varepsilon}^{234}$ indicates the same with $3$ links between Tx1 to Rx2, Rx3 and Rx4.}.

Below we discuss the position estimation performance of the Type-B architecture. Here we use all the three links from transmitter to Rx2, Rx3 and Rx4. Each CNN pipeline will carry the CIR from a single link as shown in the Fig.~\ref{fig:typeB}  with $L=3$. We use similar training and testing strategy described above for Type-A architecture. Comparing the performance between Type-A and Type-B architectures in Fig.~\ref{fig:PP_TypeABC} (green and blue plots), there is an improvement in the Type-B algorithm as it reduces the average position error from $\mu_{\varepsilon}^{234}=1.1\mb{\,m}$ to $\mu_{\varepsilon}^{234}=0.8\mb{\,m}$. However, this improved performance comes at the cost of increased computational requirements, memory usage, and training time due to the architecture's greater complexity and higher number of model parameters.

\subsubsection{Tree-Based Architectures (Type-C)}
\label{sss:tree}

\begin{figure}[t]
  \centering
  \fbox{\includegraphics[width=0.8\columnwidth]{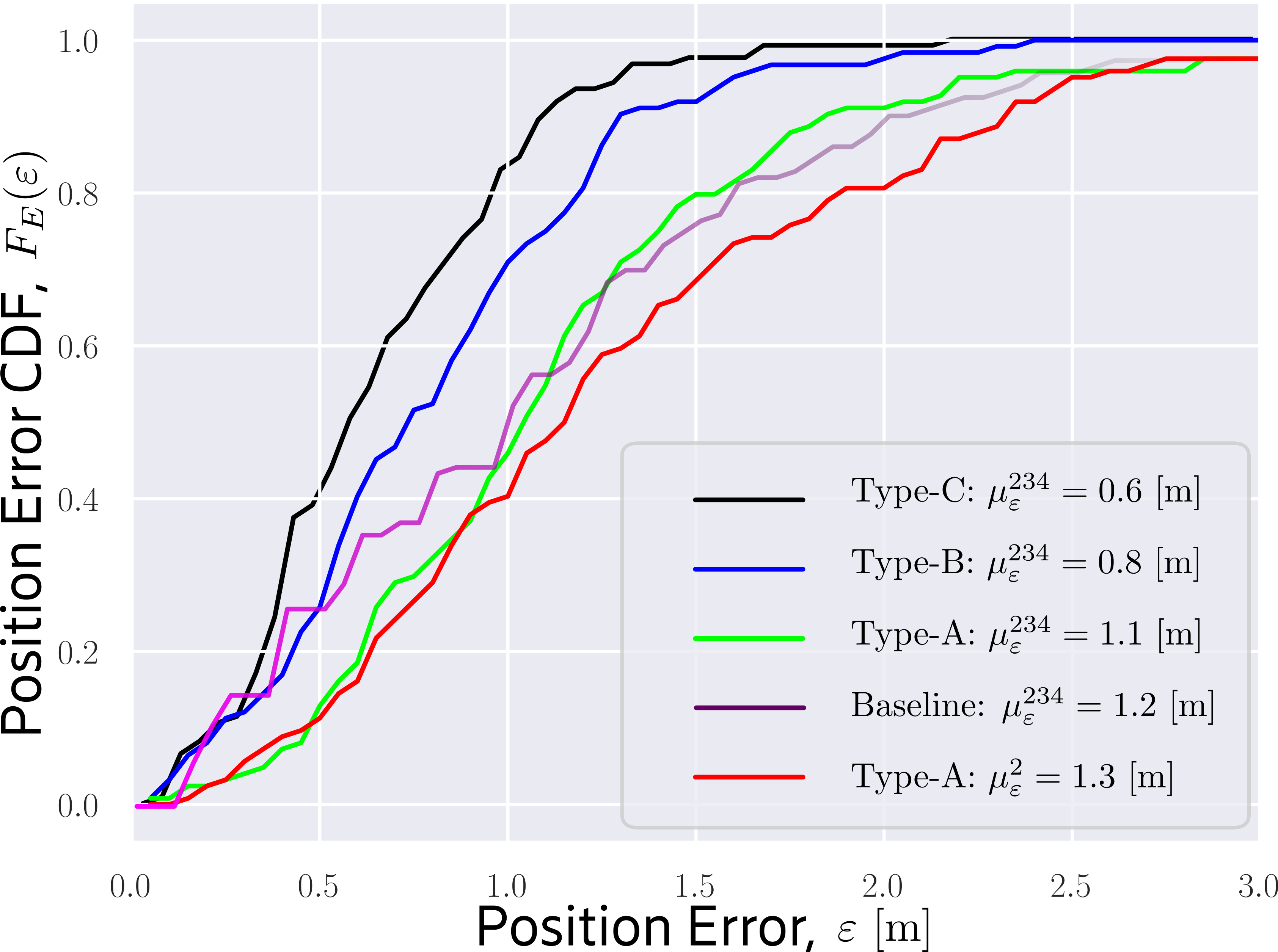}}
  \caption{Positioning performance is denoted through CDF of position estimation error ($\varepsilon$) for Type-A, B, C and baseline methods. Type-C architecture provides best positioning performance.}
  \label{fig:PP_TypeABC}
\end{figure}

In Fig.~\ref{fig:PP_TypeABC}, the positioning performance of the (hyperparameter tuned) Type-C architecture is compared with the CNN-based architectures. The Type-C architecture outperforms the Type-A and Type-B to yield the best positioning performance. Type-A and Type-B architectures are based on deep neural networks. Moving from Type-A to Type\nobreakdashes-B approximately increases the total number of model-parameters (and hence the compute and inference time) of the neural network by  a factor $L$. However, Type\nobreakdashes-C is not based on deep neural network but on tree-based ensemble architecture. Although the Type\nobreakdashes-C model has fewer parameters compared to Type-B, it may not be able to leverage hardware accelerators as effectively as deep neural networks. Therefore, the actual choice of these architectures depends on the available underneath AI hardware and accelerators.

\subsection{Baseline Evaluation}
\label{resultsBE}
Since the measurements are quite densely collected, we develop a baseline method, which uses a database of CIRs to positions mapping. For inference of new samples (test set) the Euclidean distance between the test CIR sample to all CIR measurements in the database is calculated, and the position corresponding to the minimum distance is selected. The performance is reported in Fig.~\ref{fig:PP_TypeABC}, notice that the baseline method is comparable to Type-A, while Type-B and Type-C clearly outperform the baseline.

\section{Summary and Conclusion}
In this study, measurements from a multi-static indoor deployment are considered. The main idea is to exploit the perturbation in the impulse response due to the presence or absence of the target towards sensing using CNN and ensemble based methods. From the results (refer to Fig.~\ref{fig:detectionResult}), target detection is a straightforward problem even with a single link (i.e., Tx1 to Rx2) detection accuracy of more than 90\% can be achieved. However, position estimation is a much more challenging problem requiring multiple links and more complex AI architectures to optimize performance. From the results, (refer to Fig.~\ref{fig:PP_TypeABC}), for fixed number of links, the choice of AI architecture makes a significant difference. For example, with all three links utilized, the position error reduces from $1.1\mb{\,m}$ to $0.8\mb{\,m}$ using a different CNN architecture (Type-A to Type-B) and to $0.7\mb{\,m}$ by shifting to ensemble-based architectures like XGBoost.  Also, for a fixed AI architecture, impulse response from multiple links enhances the performance. Finally, the overall results indicate that AI can provide enhanced performance in passive target localization.

\bibliography{main}

\begin{thebibliography}{1}
\providecommand{\url}[1]{#1}
\csname url@samestyle\endcsname
\providecommand{\newblock}{\relax}
\providecommand{\bibinfo}[2]{#2}
\providecommand{\BIBentrySTDinterwordspacing}{\spaceskip=0pt\relax}
\providecommand{\BIBentryALTinterwordstretchfactor}{4}
\providecommand{\BIBentryALTinterwordspacing}{\spaceskip=\fontdimen2\font plus
\BIBentryALTinterwordstretchfactor\fontdimen3\font minus
  \fontdimen4\font\relax}
\providecommand{\BIBforeignlanguage}[2]{{%
\expandafter\ifx\csname l@#1\endcsname\relax
\typeout{** WARNING: IEEEtran.bst: No hyphenation pattern has been}%
\typeout{** loaded for the language `#1'. Using the pattern for}%
\typeout{** the default language instead.}%
\else
\language=\csname l@#1\endcsname
\fi
#2}}
\providecommand{\BIBdecl}{\relax}
\BIBdecl

\bibitem{yajnanarayana2023eucnc}
V.~Yajnanarayana and H.~Wymeersch, ``Multistatic sensing of passive targets
  using {6G} cellular infrastructure,'' in \emph{2023 Joint European Conference
  on Networks and Communications \& 6G Summit (EuCNC/6G Summit)}, 2023, pp.
  132--137.

\bibitem{saleh-1987-statis-model}
\BIBentryALTinterwordspacing
A.~Saleh and R.~Valenzuela, ``A statistical model for indoor multipath
  propagation,'' \emph{IEEE Journal on Selected Areas in Communications},
  vol.~5, no.~2, pp. 128--137, 1987. [Online]. Available:
  \url{https://doi.org/10.1109/jsac.1987.1146527}
\BIBentrySTDinterwordspacing

\bibitem{mfm}
I.~P. Roberts, ``{MIMO} for {MATLAB}: A toolbox for simulating {MIMO}
  communication systems in {MATLAB},'' \url{http://mimoformatlab.com}, Jan.
  2021.

\bibitem{yajnanarayana2024-bistatic}
\BIBentryALTinterwordspacing
V.~Yajnanarayana and P.~Geuer, ``Bi-static sensing in {OFDM} wireless systems
  for indoor scenarios,'' \emph{CoRR}, 2024. [Online]. Available:
  \url{http://arxiv.org/abs/2403.04201v1}
\BIBentrySTDinterwordspacing

\bibitem{chris2023eucnc}
C.~Mollen, G.~Fodor, R.~Baldemair, J.~Huschke, and J.~Vinogradova, ``Joint
  multistatic sensing of transmitter and target in {OFDM}-based {JCAS}
  system,'' in \emph{2023 Joint European Conference on Networks and
  Communications \& 6G Summit (EuCNC/6G Summit)}, 2023, pp. 144--149.

\bibitem{chen2016-xgboos}
\BIBentryALTinterwordspacing
T.~Chen and C.~Guestrin, ``Xgboost: A scalable tree boosting system,'' in
  \emph{Proceedings of the 22nd ACM SIGKDD International Conference on
  Knowledge Discovery and Data Mining}, 8 2016, p. nil. [Online]. Available:
  \url{http://dx.doi.org/10.1145/2939672.2939785}
\BIBentrySTDinterwordspacing

\end{thebibliography}
\bibliographystyle{IEEETran}
 
\end{document}